\UseRawInputEncoding
\documentclass[reprint,aps,pre,showkeys,showpacs,superscriptaddress,twocolumn,amsmath,amssymb,longbibliography]
{revtex4-1} 
\usepackage{graphics}
\usepackage{bm}
\usepackage{subfigure}
\usepackage{epsfig}
\usepackage[bookmarks,colorlinks,citecolor=blue,linkcolor=blue]{hyperref}

\usepackage{natbib}
\usepackage{siunitx}
\usepackage{soul}
\setstcolor{red}
\usepackage{array}
\usepackage{multirow}
\usepackage{amsmath}
\usepackage{esvect}
\usepackage{pdfcomment}
\usepackage{marginnote}
\usepackage{xcolor}
\definecolor{mygreen}{HTML}{2a9d8f}
\definecolor{myyellow}{HTML}{e09f3e}
\DeclareSIUnit\Molar{\textsc{m}}

\usepackage{xcolor}
\usepackage[normalem]{ulem}

\usepackage{graphicx}	
\begin{document}
\title{Graph Neural Network for Multitask Prediction of Rheological and Microstructural Behavior in Suspensions}
\author{Armin Aminimajd}
\affiliation{Department of Macromolecular Science and Engineering, Case Western Reserve University, Cleveland, OH, 10040, USA}
\author{Joao Maia}
\email{joao.maia@case.edu}
\affiliation{Department of Macromolecular Science and Engineering, Case Western Reserve University, Cleveland, OH, 10040, USA}
\author{Abhinendra Singh}
\email{abhinendra.singh@case.edu}
\affiliation{Department of Macromolecular Science and Engineering, Case Western Reserve University, Cleveland, OH, 10040, USA}
\date{\today}

\begin{abstract}
Fast prediction of suspension rheology is fundamental for optimizing process efficiency and performance in numerous industrial settings. 
However, traditional simulations are computationally demanding due to explicit evaluation of contact networks and stress tensors in dense regimes approaching shear thickening and jamming.
%
%
This study presents a microstructure-informed multitask learning framework based on the graph neural network (GNN) that learns an implicit mapping between particle configurations and emergent microstructural and rheological properties of suspensions. 
This model simultaneously predicts particle pressure $\Pi$, viscosity $\eta$, and friction coordination $Z_\mu$, in a dynamic steady-state, without explicit knowledge of interparticle forces.
Here, semi-dilute to dense suspension systems in 2D were simulated across a wide range of shear stresses $\sigma$, spanning continuous, discontinuous shear thickening, and shear-jamming conditions.
The trained models demonstrated high correlation coefficients ($R^2$ = 0.99) with narrow mean absolute error for packing fractions up to $\phi \le \phi_J^\mu$ for all predictive targets. 
However, prediction scatter increases near jamming conditions, attributed to inherent fluctuations in suspension behavior as the critical packing fraction is approached, yet predictions remain in excellent agreement, closely following the trend of the simulated flow curves across stress evolution. 
Once trained, the model can infer rheological responses directly from structural topology, avoiding explicit stress evaluation during prediction.
The approach yields computationally efficient mesoscale surrogates for accelerated simulation with potential for real-time exploration of particulate suspension behavior.
\end{abstract}
\maketitle

\section{Introduction}
Suspensions are ubiquitous and complex materials consisting of solid particles dispersed in a fluid medium. Understanding their physics and rheological behavior is crucial for a wide array of natural, industrial, and biological systems, such as in geological transport phenomena \cite{vercruysse2017suspended}, flow control in paints \cite{eley2019applied} and food products \cite{goralchuk2019food}, controlled drug delivery \cite{townsend2019flow, kumar2022particle}, and the diagnosis of cardiovascular diseases \cite{takeishi2019haemorheology}. Under deformation, suspensions exhibit non-Newtonian behaviors, such as discontinuous shear thickening and jamming, driven by the interplay between bulk deformation and interparticle interactions. 
Recent studies reveal the key role of frictional contacts in predicting the rheological behavior of suspensions \cite{Mari_2014, Morris_2020, Clavaud_2017, Clavaud_2025, bossis2022discontinuous, Comtet_2017, sharma2025frictional}. Generally, the relative viscosity $\eta_r$ increases with the particle volume fraction $\phi$, reaching a critical point where the suspension can no longer flow as it approaches the frictional jamming limit $\phi_j^\mu$, which is influenced by the particle asperities \cite{Peters_2016, Barik_2022, Singh_2019, Singh_2020, Singh_2022, Singh_2024, Clavaud_2025}. 

However, predicting the rheological and microstructural properties of dense suspensions, especially near the jamming transition, presents significant challenges for computational approaches that arise from the complex interplay of short-range interactions, multi-scale phenomena, and the inherent non-equilibrium nature of these systems, which makes simulation extremely costly and in some cases impractical \cite{aminimajd2025robust,boodaghidizaji2022multi}. 
On the other hand, experimental methods are unable to accurately differentiate particle pressure from fluid contribution~\cite{guazzelli2018rheology}, frictional contacts \cite{guazzelli2018rheology, Pradeep_2021, Pradeep_2020}, and even viscosity due to issues such as wall-slip effects, sedimentation, and particle migration \cite{guazzelli2018rheology}. 
Therefore, efficient prediction and characterization of rheological and microstructural properties require advanced techniques to infer non-linear behaviors coupled with microstructural dynamics. 
%

%
Given the highly nonlinear physics and recent advances in machine learning approaches to soft matter~\cite{ashwin2022deep, rossi2022identification, binel2023online,Jackson_2019,Barrat_2023,Ferguson_2017,Dulaney_2021,Gartner_2019,de2023data,floryan2022data,mehdi2024enhanced}, recent years have witnessed the application of these concepts to the rheology of complex fluids~\cite{mangal2024data,mahmoudabadbozchelou2021rheology,dabiri2025detailed,lennon2023scientific,jin2023data}.
Data-centric models map input and output to uncover hidden relationships within data and often need diverse features, including material properties, processing parameters, and geometrical characteristics, to effectively capture the underlying physics and rheological properties \cite{barcelos2022coupling, boodaghidizaji2022multi}. 
These models, such as Support Vector Machines (SVM) and K-Nearest Neighbors (KNN), have been integrated with microfluidic devices for in situ viscosity measurement \cite{mustafa2023machine}, and the Convolutional Neural Network (CNN) has been used to predict the rheological properties of Bingham fluids using stereo camera images \cite{ponick2022image}.
Recent studies have utilized fractional Physics-Informed Neural Networks (fPINNs) to solve fractional differential equations in the constitutive modeling of complex materials, such as viscoelastic fluids and soft matter systems \cite{dabiri2025detailed}.
In addition, fPINNs have been employed to capture the time evolution of shear stress responses and material parameters \cite{mahmoudabadbozchelou2021rheology}, learning of particle stress development \cite{howard2023machine}, and inferring steady-state shear viscosity from velocity and pressure data in time-dependent 3D flows \cite{reyes2021learning}.
Despite these advances, the physics-informed approach can be constrained by the particular model parameters, flow protocols \cite{mangal2024data}, and boundary conditions \cite{sheikholeslami2025physics, wang2024machine}. Even with embedded physics, this approach requires model retraining to accommodate varying conditions. Some studies utilized physics-compatible neural operators to generalize models across various problem settings in complex fluid systems \cite{azizzadenesheli2024neural, saberi2025rheoformer} without retraining. However, the application to history-dependent and highly nonlinear flows remains limited. Given this, machine learning approaches to predict rheological and microstructural properties near transitions, such as discontinuous shear thickening (DST) and shear jamming (SJ), remain challenging, where physics is inherently fluctuating and noisy.

Here, we utilize microstructural features to infer the micro- and macro-behavior of suspensions, particularly during dramatic transitions such as shear thickening and jamming. 
In this study, supervised Multitask learning (MTL) is employed, which typically yields better performance and robustness compared to models trained solely on a single property. 
This gain is mainly because it reduces the risk of overfitting any single property by leveraging shared parameters and inherent correlations across tasks \cite{caruana1997multitask}.
This effectively means that the information learned from one property can help the model to generalize to predict other properties \cite{tang2024multi}. 
Multitasking assumes a relationship between different prediction tasks, allowing the model to exploit shared information and the underlying correlations between properties~\cite{kuenneth2021polymer}; therefore, it builds an improved latent representation or graph embedding~\cite{roy2025materials, gupta2025benchmarking}. 
Moreover, MTL can achieve the target accuracy at a lower computational cost than single-task approaches \cite{roy2025materials}.
Here, we predict three distinct and correlated properties that are crucial for dense suspension flow: (1) relative viscosity $\eta_r$, (2) particle pressure $\Pi$, and (3) frictional coordination number $Z_\mu$, in the packing fraction range where suspension exhibits strong CST, DST, and DST-SJ behaviors. 


This bottom-up method involves generating necessary datasets through simulation, extracting microscopic information for a single snapshot in a dynamic steady-state at each shear stress, such as relative distances and dimensionless gaps, and using them to build graphs for further use in multitask prediction of macroscopic properties via the DeepGCN approach. 
Here, the values of $\eta_r$, $\Pi$, and $Z_\mu$ are likewise extracted for a snapshot in dynamic steady-state for an imposed shear stress, rather than the traditional approach where an average (over steady-state) is reported. Then, the predicted macroscopic properties emerge directly from interparticle interactions encoded in their geometric features, allowing us to understand how structure determines function. 
In previous studies, we demonstrated the efficacy of this approach in accurately predicting frictional contact networks (FCN) across various conditions, including changing processing parameters to particle physical constraints without explicit knowledge of interparticle forces \cite{aminimajd2025scalability, aminimajd2025robust}. 
A similar approach is applied in this study to infer bulk and microstructural properties from limited input features, using only physical intuition and a data-driven approach that does not rely on precomputed heuristic features. 
This reliance on minimal input enables the models to be generalizable and symmetry-aware, reducing their susceptibility to overfitting. Finally, the results indicate excellent agreement between simulation and predicted values, uncovering the potential of this approach to bypass complex computations. 

The current study is organized as follows: Sec.~\ref{Sec:Methods} provides detailed information about the simulation and machine learning methods; Sec.~\ref{Sec:Results} compares the direct simulation results and the MTL architecture that we present here. Finally, Sec .~\ref {Sec:Conclusion} provides concluding remarks and an outlook for future directions in the application of these approaches in dense suspension rheology.

%
\begin{figure*}
    \includegraphics[trim = 10mm 10mm 10mm 0mm, clip,width=1\textwidth,page=1]{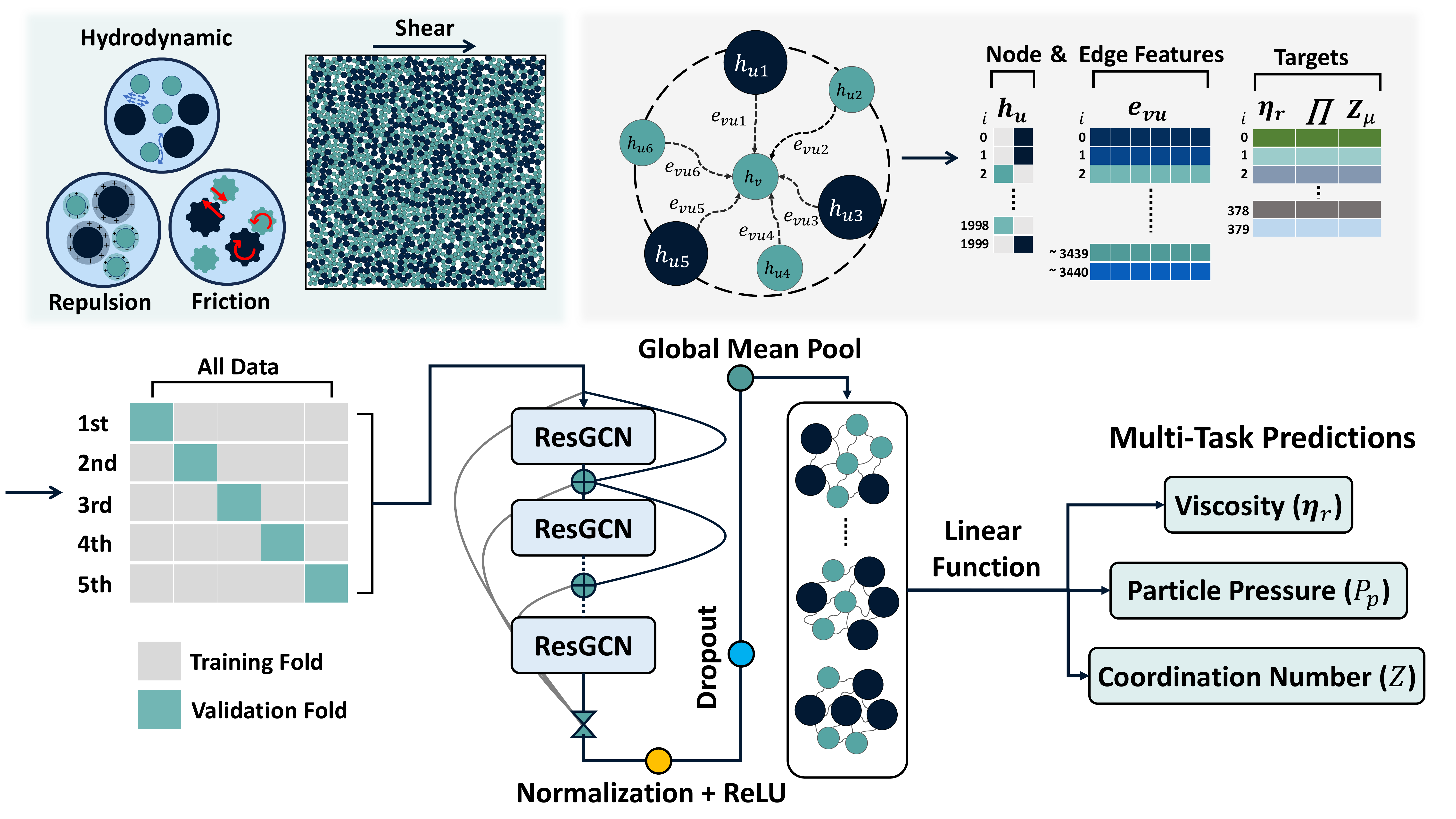}
    \caption{\textbf{Schematic illustration of the machine learning methodology.} Schematic for the simulation, training process and multitask prediction of viscosity $\eta$, particle pressure $\Pi$, and frictional coordination number $Z_\mu$ employing Deep Graph Convolutional Neural Network (DeepGCN). Simulations were conducted through the LF-DEM method to generate configurations, which were then transformed into graphs for input into the graph neural network (GNN), where particle features and the interparticle interactions are represented by node features and edge attributes, respectively. The datasets consist of particle radius as node feature ($h_u$) and relative distance between particles ($R_\text{ij}$), x and y components of $R_\text{ij}$, sine and cosine of $R_\text{ij}$ between particles as edge attributes ($e_\text{vu}$). Next, the datasets were divided into 5 folds, where at each iteration, four folds were used for training and the remaining fold was reserved for validation. Then, the datasets are processed into the residual graph convolution layers (ResGCN) to update node features through exchanging information between nodes and their neighboring edges for the specific particle. The output of each layer was normalized and was subjected to a non-linear activation function (ReLU) to improve the complexity and capacity of the model to capture hidden interparticle relationships. This process is repeated for the number of layers, and finally, a linear layer predicts the properties of each graph through three outputs. } 
    \label{fig:fig1}
\end{figure*}

\begin{figure*}
    \centering
    \includegraphics[trim = 0mm 140mm 450mm 0mm, clip,width=0.90\textwidth,page=2]{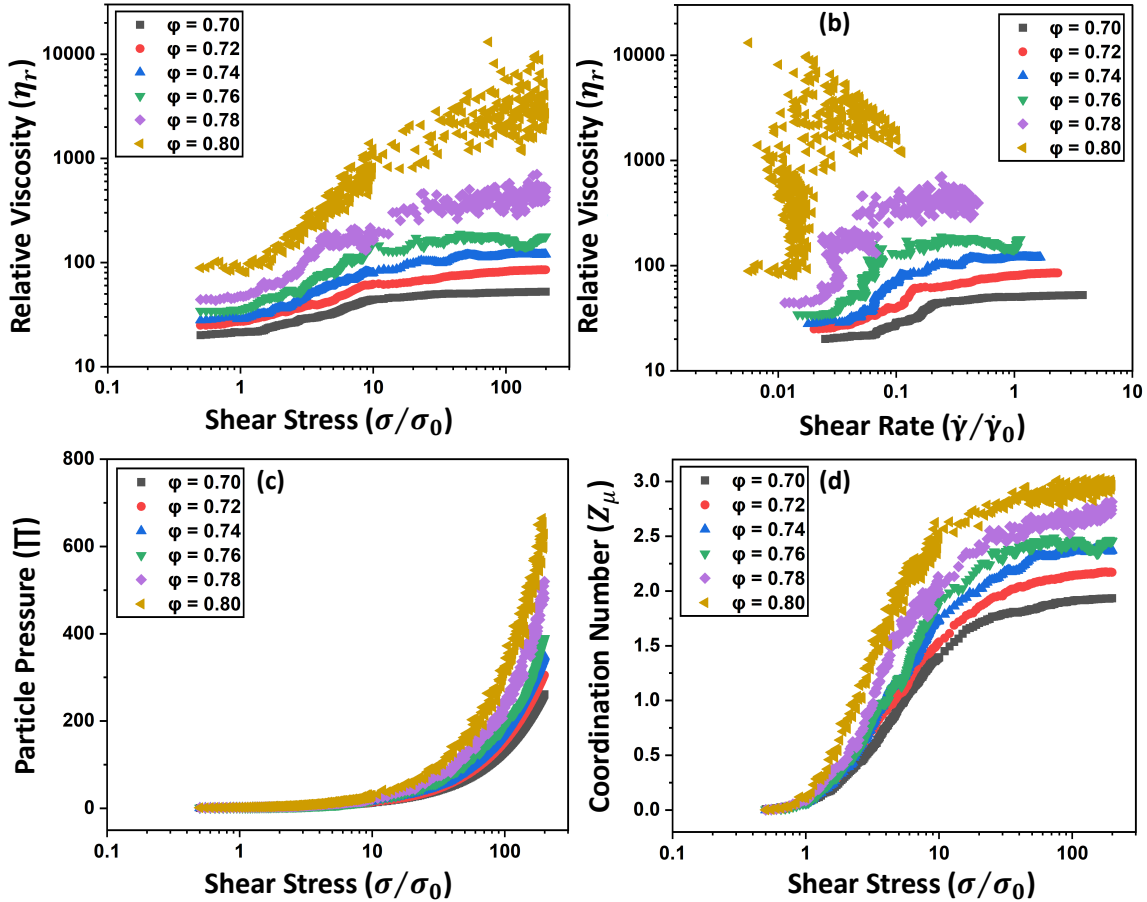}
    \caption{\textbf{Rheological and microstructural response under shear.} Rheological and microstructural behavior of suspensions for packing fractions $\phi = 0.70-0.80$. (a) Viscosity $\eta$ as a function of shear stress $\sigma_\text{xy}$, (b) viscosity $\eta$ vs shear rate $\dot{\gamma}$, (c) particle pressure $\Pi$, and (d) frictional coordination number $Z_\mu$ vs $\sigma_\text{xy}$.}
    \label{fig:fig2}
\end{figure*}

\section{Methods}
\label{Sec:Methods}
\paragraph*{{Simulating suspensions}:}
In this study, Lubrication Flow-Discrete Element Method (LF-DEM) is used to simulate semi-dilute to dense suspensions ($\phi \in [0.70 - 0.80]$) with applied simple shear flow of non-Brownian frictional particles suspended in a Newtonian fluid~\cite{Morris_2020, Mari_2015, Singh_2020}. 
Previously, we have demonstrated that 2D and 3D exhibit similar rheological behavior when the packing fraction is scaled with the frictional jamming point $\phi_J^\mu$~\cite{aminimajd2025scalability,Gameiro_2020}.
Given this, for simplicity and computational efficiency, we simulated 2D suspensions to generate high-fidelity data to feed our data-driven approach. Here, the system follows Stokes flow with hydrodynamic interactions $\vec{F}_{\mathrm{H}}$ and contact forces $\vec{F}_{\mathrm{C}}$ governing particle dynamics and macroscopic rheology:
\begin{equation}\label{eq:1}
    \vec{0} = \vec{F}_{\mathrm{H}}(\vec{X},\vec{U}) + \vec{F}_{\mathrm{C}}(\vec{X})~,
\end{equation}
where $\vec{X}$ and $\vec{U}$ represent the position and velocity of the particles, respectively. 
The hydrodynamic force calculations account for both two-body lubrication interactions and one-body Stokes drag~\cite{Seto_2013a, Mari_2014}. The dominant component of the lubrication forces exhibits a divergence proportional to $1/h$, where $h$ denotes the surface-to-surface separation between particles. Consistent with prior studies~\cite{Melrose_1995, Morris_2020, Seto_2013a, Singh_2018, Singh_2020, Singh_2022}, it is presumed that lubrication forces break down below $h_{\mathrm{min}}/a = 0.001$, where $a$ refers to the radius of the smaller particle. To accommodate this, lubrication forces are regularized below $h_{\mathrm{min}}$, thereby allowing direct particle contacts.
For modeling contact forces, the conventional Cundall \& Strack framework~\cite{Cundall_1979} is employed using Luding's algorithm~\cite {Luding_2008}. In this approach, contact forces are implemented using linear springs, which are activated only for overlap  $\delta^\text{(i,j)} = a_i + a_j - |r_i - r_j| > 0$.

For a given particle pair ($i,j$) with overlap $\delta$ and a unit vector $\boldsymbol{n}$ connecting their centers, normal contact $\boldsymbol{F}_{\mathrm{C,n}}$, sliding friction force $\boldsymbol{F}_{\mathrm{C,t}}$ and sliding friction torque $\boldsymbol{T}_{\mathrm{C,t}}$ are calculated using the following: 
\begin{subequations}
\begin{equation}
\boldsymbol{F}_{\mathrm{C,N}}^{(i,j)} = k_{n}\delta^{(i,j)}\boldsymbol{n}_{ij}~,
\end{equation}
\begin{equation}
\boldsymbol{F}_{\mathrm{C,T}}^{(i,j)} = k_{t} \boldsymbol{\xi}^{(i,j)}~,
\end{equation}
\begin{equation}
\boldsymbol{T}_{\mathrm{C,T}}^{(i,j)} = a_i \boldsymbol{n}_{ij} 
\times \boldsymbol{F}_{\mathrm{C,T}}^{(i,j)}~.
\end{equation}
\end{subequations}
%
%
The unit vector $\boldsymbol{n_{ij}} \equiv (\boldsymbol{r}_i - \boldsymbol{r}_j)/|\boldsymbol{r}_i - \boldsymbol{r}_j|$ points from the particle $j$ to $i$, with $a_{ij} \equiv 2 a_i a_j/(a_i + a_j)$ referring to the reduced radius.
Here, $k_n$ and $k_t$ denote spring constants for the normal and sliding contact forces, respectively. Sliding frictions obey Coulomb's friction laws, i.e., $|\boldsymbol{F}_{\mathrm{C,T}}^{(i,j)}| \le \mu_s |\boldsymbol{F}_{\mathrm{C,N}}^{(i,j)}|$ where $\mu_s$ is sliding friction coefficient.
Finally, the total contact force and torque are calculated as follows:
\begin{subequations}
\begin{equation}
\boldsymbol{F}_{\mathrm{C}}^{(i,j)} = 
\boldsymbol{F}_{\mathrm{C,nor}}^{(i,j)}  + \boldsymbol{F}_{\mathrm{C,slid}}^{(i,j)}~,
\end{equation}
\begin{equation}
\boldsymbol{T}_{\mathrm{C}}^{(i,j)} = 
a_i \boldsymbol{n}_{ij} \times \boldsymbol{F}_{\mathrm{C,slid}}~.
\end{equation}
\end{subequations}
The Critical Load Model (CLM) accounts for rate dependence by requiring the normal force to surpass a critical threshold, $F_0$, to activate interparticle friction $\mu_s$. This threshold force establishes a characteristic stress scale, $\sigma_0 = F_0/6\pi a^2$, that non-dimensionalizes the shear stress.

We employ Lees-Edwards periodic boundary conditions with an assembly of $N=2000$ (and $N=400$ to examine the size dependency on prediction accuracy) non-Brownian bidisperse particles within a unit cell. To prevent ordering, an equal volume of small particles (radius $a$) and large particles (radius $1.4a$) is mixed in the systems that reproduce the features of experimentally measured rheological responses in dense suspensions~\cite{Mari_2014, Singh_2020, Singh_2022}.

For the model training purpose, suspension configurations are generated by varying shear stress $\sigma/\sigma_0 \in [0.5, 200]$ and packing fraction $\phi \in [0.70, 0.80]$ with particle friction constant fixed to $\mu = 0.5$. Given the interest in the packing fractions close to DST and SJ conditions, the simulations are performed at controlled shear stress~\cite{Singh_2018, Mari_2015}. With this, the shear rate $\dot{\gamma}(t)$ fluctuates over time, which is averaged over the steady state. Here, relative viscosity is computed as $\eta_r(t) = \sigma/\eta_0\dot{\gamma}(t)$, with $\eta_0$ being the viscosity of the suspension liquid, particle pressure $\Pi$, is defined as $\Pi \equiv - (\Sigma_\text{11} + \Sigma_\text{22} + \Sigma_\text{33})/3$, and the frictional coordination number $Z_\mu$ is calculated as the average number of frictional contacts. 

\paragraph*{{Machine Learning Method:}}
Neural networks often suffer from the vanishing gradient problem, leading to error accumulation in deep architectures. To mitigate this issue, a variant of the graph neural network (GNN) known as the Deep Graph Convolutional Neural Network (DeepGCN) is employed, which enables a deep and robust learning procedure by utilizing a residual connection~\cite{Guohao_2020, li_2019deepgcns}. This architecture has demonstrated accurate prediction of frictional contact network in our previous studies \cite{aminimajd2025scalability, aminimajd2025robust}.

The schematic for our machine learning approach is depicted in Fig. \ref{fig:fig1}. The first step involves simulating suspensions with simple shear using the LF-DEM method in a stress-controlled manner at fixed packing fractions. The dataset consists of 380 graphs covering a broad range of shear stresses, $\sigma/\sigma_0 = 0.5 - 200$, for each packing fraction, $\phi = 0.70 - 0.80$. The suspensions naturally represent graphs, so here, particles and interparticle interactions can be represented as nodes and edges. Given this, the generated configurations are converted into graphs, which are used to feed the GNN. Here, we used geometric features as input, where (1) the particle radius is considered as the node feature, which is one-hot encoded (0 or 1) to represent the bidisperse system; (2) edge attributes contain dimensionless gap, relative distance between particles ($R_{ij}$), $x$ and $y$ components of $R_{ij}$, sine and cosine of $R_{ij}$ corresponding to deformation direction. The choice of these inputs provides a consistent, smooth training procedure that relies solely on the particles' physical positions.
This feature selection ensures efficient future characterization of suspensions by providing easy-to-generate data, either through simulation or experiments, using a simple snapshot. After creating the graphs, each will be assigned three target values representing viscosity $\eta_r$, particle pressure $\Pi$, and frictional coordination number $Z_\mu$, which have been extracted from each snapshot rather than averaging steady state values.

Here, $K$-fold cross-validation is used to ensure obtaining an unbiased estimate of the model performance on unseen data and to mitigate the risk of overfitting. 
In this study, the dataset is divided into five folds, with one fold (20\% of the dataset) reserved for testing and the remaining four used for training (80\% of the dataset). This process is repeated five times, each with different random weight initializations and a different fold as the test set. 
The datasets were selected at random to improve the robustness of the training process and ensure a diverse data distribution. The dataset is then passed to the DeepGCN training pipeline for model training in each fold. 
The training process involves taking information into the network consisting of $N_l$ layers with Residual layers (ResGCN), where nodes are updated through message passing processing, i.e., sending and receiving information to neighboring nodes for a specific node through edges:

\begin{equation}
    h_v^{l} =  h_v^{(l-1)} + \sum_{u \in \mathcal{N}(v)} f(h_u^{(l-1)}, h_v^{(l-1)}, e_{uv})
    \label{eq:3}
\end{equation}

Here, $h_v^{(l)}$ represents the updated feature of node $v$ at layer $l$, the set $ \mathcal{N}(v) $ denotes the neighbors of node $v$, i.e., all nodes directly connected to it, the function $f$ operates on the features of both node $v$ and its neighbors $u \in \mathcal{N}(v)$ from the previous layer $l-1$, along with the edge feature $e_\text{uv}$ between them. This function typically involves concatenating node and edge features, followed by a non-linear aggregation. The result of this aggregation is then added to the original node feature $h_v ^l$ through a residual connection. This residual design helps mitigate vanishing gradient issues, thereby enabling more effective training of deep graph networks.
For the graph regression task, each input graph is treated as a single data point, and its representation is updated through ResGCN. At each layer, features are transformed using learnable weights and biases. The outputs are batch-normalized and passed through a non-linear activation function to enhance the model's capacity to capture complex interparticle interactions and hidden structural relationships. Here, the rectified linear unit (ReLU) was chosen as the activation function due to its efficiency by deactivating neurons with negative outputs during forward propagation \cite{nair2010rectified}.
Finally, a linear layer combines processed features into three outputs to predict the properties of each graph. Using this GNN approach, we will demonstrate that the assembly of physical information encoded in nodes and edges is reflected in the system’s rheological and structural behavior and thus an accurate prediction of their corresponding properties. 

During training, the disparity between predicted and ground truth values is evaluated using the Huber loss (or \texttt{SmoothL1Loss}) by the backpropagation method. This loss function combines properties of L1 loss (Mean Absolute Error) and L2 loss (Mean Squared Error) and ensures a balance between them, providing robustness to outliers while maintaining smoothness around zero. It creates a criterion that uses L2 Loss if the absolute element-wise error falls below the beta value ( = 1, threshold parameter) and an L1 term otherwise:

\begin{equation}
    Loss(\hat{y}, y) = 
    \begin{cases} 
        0.5 \times (\hat{y} - y)^2 & \text{if } |\hat{y} - y| < 1 \\ 
        |\hat{y} - y| - 0.5 & \text{otherwise}
    \end{cases}
\end{equation}

Where: $\hat{y}$ and y are the predicted and target values, respectively.
Adam optimizer is used, which is an adaptive gradient-based optimization method to minimize the loss value (learning rate, $\alpha = 5 \times 10^3$) \cite{kingma2014adam}.
The maximum number of training iterations is set to 5000, with an early stopping criterion that stops the training process if the loss does not improve for a specific consecutive epochs. The model parameters corresponding to the lowest loss are saved for future predictions on unseen datasets.
Before training, the relative viscosity $\eta_r$ and particle pressure $\Pi$ are scaled logarithmically to bring their values into a comparable range. The frictional coordination number is kept in its original form. In this multitask learning setup, all targets share a common network, and their loss values are computed as follows:

\begin{equation}
    l_\text{total} = l_\eta + l_\text{P} + l_Z
    \label{eq:4}
\end{equation}

The mean absolute error (MSE) and coefficient of determination $R^2$ are used as metrics to evaluate the performance of multitask regression models.
The model's hyperparameters are selected based on optimal performance on a validation set, comprising configurations not present in the training set. Subsequently, we assess the model's performance on a separate and independent test set. 

To save computational time and reduce the number of parameters, a shallow neural network consisting of two hidden layers with 64 neurons is selected. These hyperparameters were kept constant for all trained models for computational time management and performance comparison. In such a shallow network, the nodes seek their immediate neighbors to pass information and update themselves, which turned out to be sufficient to map underlying latent features in the suspension system of this study's interest. 

\begin{figure*}
    \centering
    \includegraphics[trim = 0mm 200mm 145mm 0mm, clip,width=0.95\textwidth,page=3]{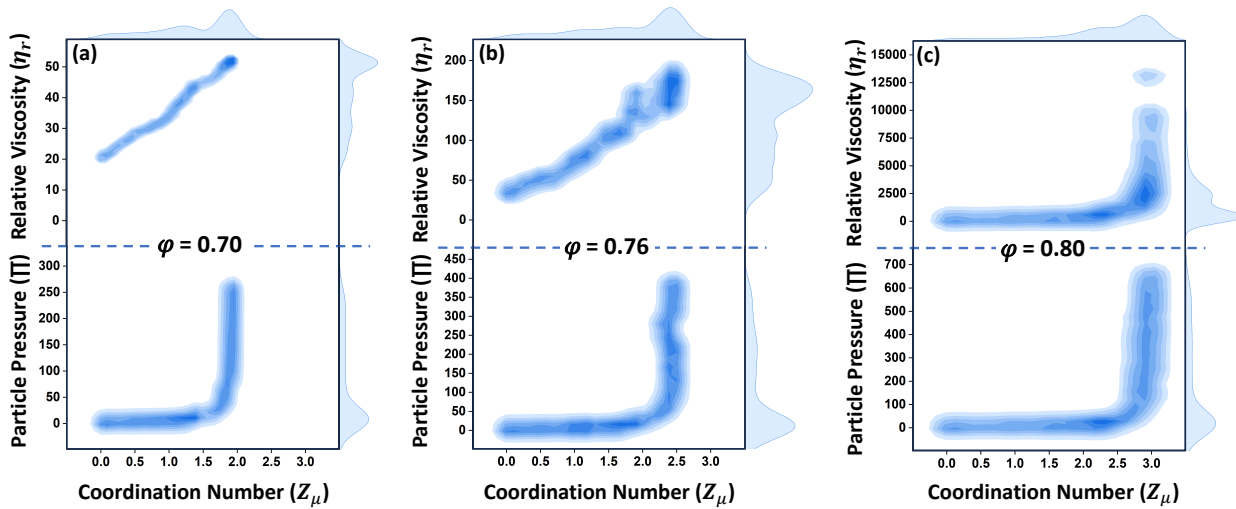}
    \caption{\textbf{Heterogeneity in simulated data sets.} Density plots and marginal distributions to visualize the heterogeneity of the datasets and the relationship of the targets used to train the models. The viscosity $\eta$, particle pressure $\Pi$ are plotted as a function of frictional coordination number $Z_\mu$ for (a) $\phi =$ 0.70, (b) 0.76, and (c) 0.80.}
    \label{fig:fig3}
\end{figure*}
\begin{figure*}
    \includegraphics[trim = 0mm 20mm 420mm 0mm, clip,width=1\textwidth,page=4]{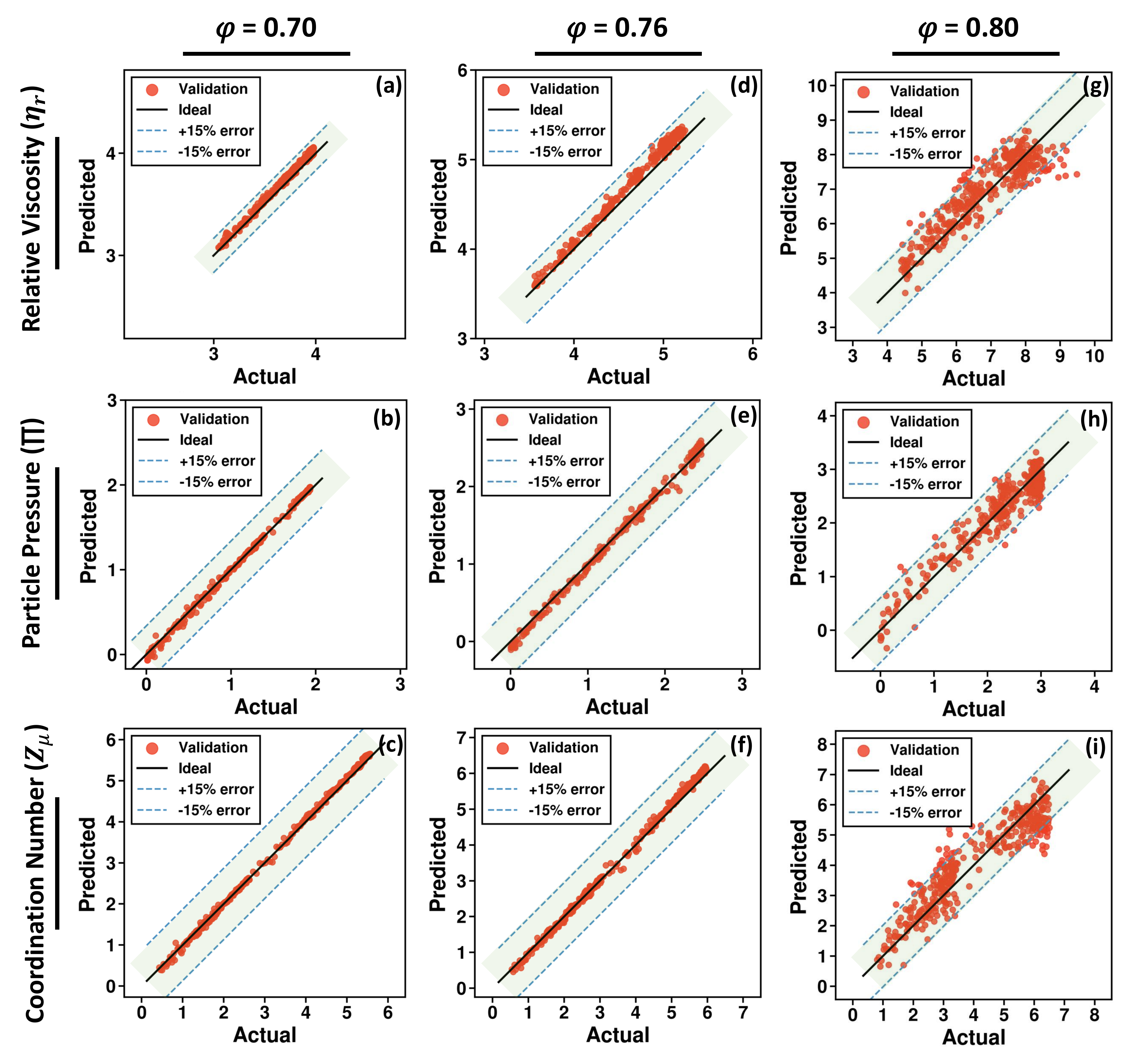}
    \caption{\textbf{Comparison between simulations and machine learning predictions.} Actual vs predictions of relative viscosity $\eta_r$, particle pressure $\Pi$, and frictional coordination number $Z_\mu$ by multitask regression based on DeepGCN for packing fractions (a-c) $\phi = 0.70$, (d-f) 0.76, and (g-i) 0.80. The red markers depict tested examples, the solid black line and the blue dashed lines represent the ideal, and $\pm15\%$ error margin. 
    } 
    \label{fig:fig4}
\end{figure*}

\begin{figure*}
    \centering
    \includegraphics[trim = 20mm 90mm 130mm 0mm, clip,width=0.9\textwidth,page=5]{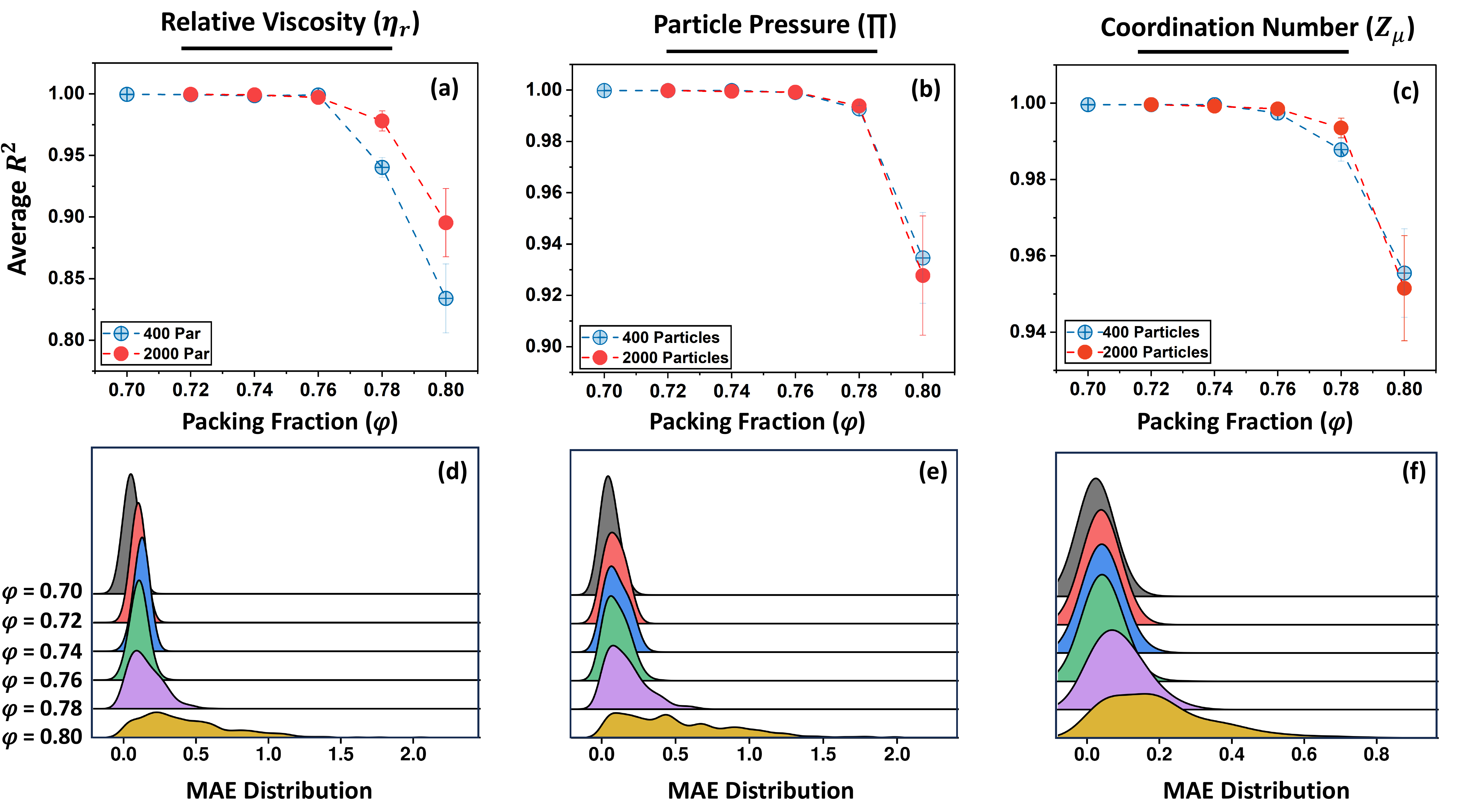}
    \caption{\textbf{Multitask learning performance.} Performance of multitask learning using DeepGCN with averaged correlation coefficient $R^2$ and mean absolute error (MAE) distribution between actual and predicted value for relative viscosity $\eta_r$ (a and d), particle pressure $\Pi$ (b and e), and frictional coordination number $Z_\mu$ (c and f) for packing fractions $\phi =$ (a-b) 0.70, (d-f) 0.76, and (g-i) 0.80.
    }
    \label{fig:fig5}
\end{figure*}

\begin{figure*}
    \includegraphics[trim = 0mm 40mm 440mm 0mm, clip,width=0.85\textwidth,page=6]{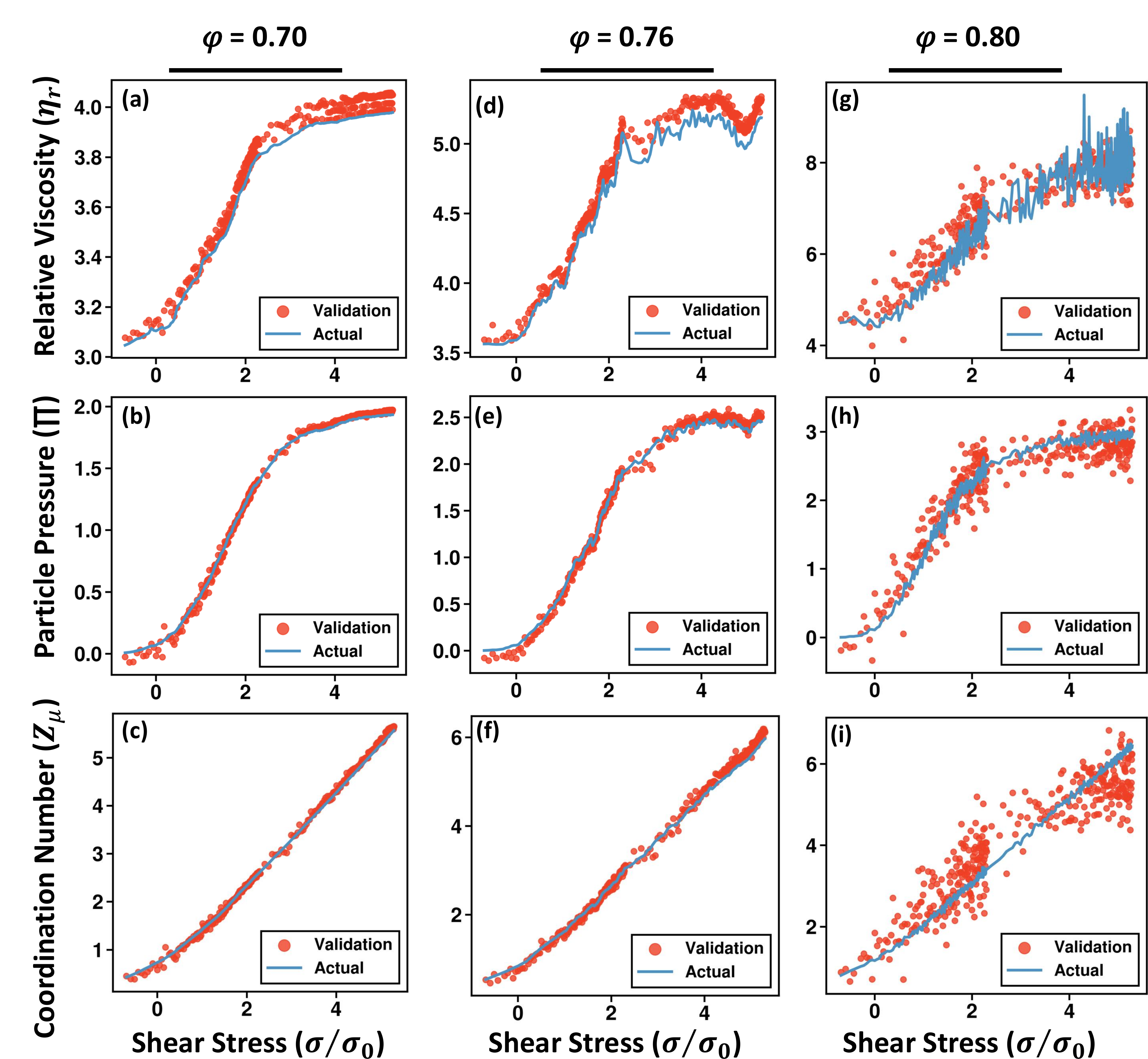}
    \caption{\textbf{Multitask learning performance.} Fitting agreement between actual and validation values predicted by multitask regression based on DeepGCN for viscosity $\eta$, particle pressure $\Pi$, and frictional coordination number $Z_\mu$ as a function of shear stress $\sigma_\text{xy}$ for $\phi = 0.70, 0.76, 0.80$.} 
    \label{fig:fig6}
\end{figure*}

%

\section{Results}
\label{Sec:Results}
Figure \ref{fig:fig2} illustrates rheological and microstructural properties of suspensions investigated in this study. Here, each point represents the result for a single snapshot at a specific shear stress. Figure~\ref{fig:fig2}a displays relative viscosity $\eta_r$ as a function of applied shear stress $\sigma/\sigma_0$, where different symbols represent different packing fractions $\phi$. We observe shear-thickening behavior, consistent with previous experimental and simulation results~\cite{Singh_2018, Singh_2020, Guy_2015, Royer_2016}. 
Here, the onset stress for shear thickening is often independent of the packing fraction, and thickening occurs between two quasi-Newtonian plateaus~\cite{Clavaud_2025}.  At low shear stress $\sigma/\sigma_0 \ll 1$, interparticle interactions are dominated by lubrication interactions with viscosity diverging at the random close packing (RCP) $\phi_J^0$. On the other hand, for $\sigma/\sigma_0 \gg 1$, almost all contacts are frictional, thus the viscosity diverges at a packing fraction $\phi_J^\mu < \phi_J^0$. The crossover between these quasi-Newtonian plateaus $(1\le\sigma/\sigma_0\le 50)$ leads to shear thickening, the extent of which depends on the packing fraction, more specifically, the distance from the frictional jamming point $(\phi_J^\mu-\phi)$. Figure~\ref{fig:fig2} (b) shows $\eta_r$ plotted as a function of dimensionless shear rate $\dot{\gamma}/\dot{\gamma}_0$. We observe continuous shear thickening for $\phi \le 0.78$. While for $\phi=0.8$, the $\eta_r(\dot{\gamma}/\dot{\gamma}_0)$ flow-curve becomes non-monotonic, showing an S-shaped curve, which is typical of the discontinuous shear thickening (DST)~\cite{Mari_2015, Singh_2018}. Another striking observation is the noise in our flow-curves, which is essentially absent for $\phi \le 0.78$. On the other hand, the simulation results for $\phi=0.8$ are highly fluctuating, consistent with previous experimental observations by Hermes et al.~\cite{Hermes_2016}. It is important to emphasize that the vicinity to the jamming point $\phi_J^\mu$, as probed here, is dynamic in nature. The shear-jammed state is stable only under shear, where a percolating (rigid) frictional contact network is inherently dynamic. It forms, might break down (then reform), or rearrange as the packing is constantly deformed under a fixed applied shear stress~\cite{Edens_2021, Gameiro_2020, Naald_2024}.

Next, we focus on particle pressure $\Pi (\sigma)$. Shown in Fig.~\ref{fig:fig2}(c) is particle pressure as a function of applied stress ($\Pi (\sigma)$) for various packing fractions $\phi$. We observe that particle pressure is zero at low stress because there are no frictional contacts between particles. As $\sigma/\sigma_0$ is increased, $\Pi$ increases roughly beyond the onset stress for shear thickening $(\sigma/\sigma_0\approx 1)$; this increase becomes much sharper at high stresses. With an increase in $\phi$, the particle pressure $\Pi$ becomes larger than the shear stress $\sigma$, and for $\phi$ approaching $\phi_J^\mu$, as deduced previously in three-dimensional simulations~\cite{Singh_2018} and a modeling based on particle migration by Morris and Boulay~\cite{Morris_1999}. This primarily implies that the bulk friction $\mu_{\text{bulk}} \equiv \sigma/\Pi<1$ as jamming packing fraction is approached, consistent with the experimental observations by Boyer et al.~\cite{Boyer_2011}.

%

Finally, we focus on the frictional coordination number $Z_\mu(\sigma)$ as plotted in Fig.~\ref{fig:fig2}(d). Here, different symbols represent simulation data for various packing fractions $\phi$. Since friction in our simulations is stress-activated, we observe that $Z_\mu=0$ for stresses $\sigma/\sigma_0 \le 1$, increases with stress, and saturates beyond $\sigma/\sigma_0 \approx 50$. The observed behavior provides the microstructural underpinning of shear thickening, as shown in Fig.~\ref{fig:fig2}(a-b). Here, a higher steady-state average $Z_\mu$ indicates more load-bearing contacts to withstand external stress. At high particle volume fractions, contact interactions between particles become dominant, which directly contribute to the particle pressure \cite{guazzelli2018rheology}.


Figure~\ref{fig:fig3} illustrates density plots and marginal distributions of rheological and microstructural properties for suspensions for $\phi = 0.70, 0.76, 0.80$ to visualize the heterogeneity (complexity) and interrelationship of the targets. Here, relative viscosity $(\eta_r)$ (top) and particle pressure $(\Pi)$ (bottom) are plotted as functions of the frictional coordination number $Z_\mu$, with color intensity indicating the density of data points. We also plot the probabilities outside on the y1 and x1 axes. As observed, the relationships among $Z_\mu$, $\eta_r$, and $\Pi$ are deeply intertwined and often non-linear, driven by the microscopic interactions and structural rearrangements within the suspension as it is sheared. We observe that both relative viscosity and particle pressure diverge as $Z_\mu$ approaches the isostatic condition $Z_\mu \approx 3$. Given the intricate relationships among these properties and the complexity and constraints of predicting them using experimental or computational approaches, we propose a straightforward method to accurately predict the rheological and microscopic behavior of dense suspensions using a multitask learning framework based on geometric features of particles for each value of $(\phi,\sigma)$. We train separate models for each packing fraction, $\phi = 0.70 - 0.80$, using datasets obtained as a function of shear stress.



In the following sections, the results of predictions for viscosity, particle pressure, and coordination number are presented as a function of shear stress for different packing fractions. Figure \ref{fig:fig4} illustrates ground truth from simulation vs predicted values by GNN for the systems consisting of $N=2000$ particles. Here, the red marks indicate the unseen datasets, which comprise the test folds from a five-fold split, totaling 380 test examples.
The solid (black) line and the dashed (blue) lines represent the ideal, and $\pm 15\%$ error.
As shown in Fig. \ref{fig:fig4}a-f, all the predictions for unseen examples fall within the $\pm 15\%$ error bounds for the dilute ($\phi = 0.70$, Fig. \ref{fig:fig4}a-c) and semi-dilute systems ($\phi = 0.76$, Fig. \ref{fig:fig4}d-f), indicating highly accurate predictions of $\eta_r$, $\Pi$, and $Z_\mu$. For the densest system ($\phi = 0.80$), although some predictions deviate, the majority of predictions still fall within the specified error margins as depicted in Fig. \ref{fig:fig4}g-i. As previously discussed, these outliers are likely due to the highly unstable conditions near the jamming transition (as shown in Fig.~\ref{fig:fig2}a-b.

Figures \ref{fig:fig5}a-c provide averaged $R^2$ results for two small and large systems containing 400 and 2000 particles as a function of $\phi = 0.70 - 0.80$, representing the effect of the system size on prediction accuracy. 
We observe that the $R^2$ values for particle pressure $\Pi$ and frictional coordination number $Z_\mu$ are nearly identical across all values of $\phi$ for the two systems (Fig. \ref{fig:fig5}b and c); thus, the effect of system size is minimal. Though the error bar (standard deviation) for $\phi=0.80$ is smaller for the system with $N=2000$ as compared to $N=400$ particles. In contrast, predictions for viscosity $\eta_r$ show strong dependence on system size, especially for near-jamming conditions, i.e., $\phi=0.78,0.8$. Generally, increasing the number of particles tends to improve the accuracy and statistical relevance of simulation results by mitigating finite-size effects and reducing biases. Thus, it is natural for a larger system to achieve higher simulation and model-training accuracies, particularly for predicting viscosity near jamming conditions. As shown, $R^2 \gtrsim 99\%$ was achieved for $\phi \lesssim 0.76$ for all targets. Nevertheless, as the suspension approaches the jamming point ($\phi = 0.80$), the accuracy decreases, yielding average $R^2$ values of 0.90, 0.93, and 0.95 for $\eta_r$, $\Pi$, and $Z_\mu$, respectively. 
Based on this, henceforth, we present results based on the model trained and tested with the system for $N=2000$ particles.
Figures \ref{fig:fig5}d–f depict the Mean Absolute Error (MAE) distribution for predictions of $\eta_r$, $\Pi$, and $Z_\mu$.
To visualize the distribution of MAE, we estimated the probability density using a Gaussian kernel density estimator with a bandwidth of 0.05.
The results indicate that the MAE distribution is sharper for semi-dilute systems ($\phi \lesssim 0.76$) but broadens as the system approaches the dense limit ($\phi \gtrsim 0.78$), reflecting increased deviations from the ground truth due to large fluctuations near the jamming transition.

Finally, Fig.~\ref{fig:fig6} presents the comparison between actual (blue line) and predicted values (red marker) as a function of shear stress for viscosity $\eta_r$, particle pressure $\Pi$, and coordination number $Z_\mu$ for $\phi = 0.70, 0.76, 0.80$. 
As shown in Figs. \ref{fig:fig6} (a-c): at $\phi = 0.70$, the validation predictions closely match the actual data, both converging to a plateau across all stress values.
At intermediate concentration, $\phi =$ 0.76, the validation still captures the overall trend for particle pressure and coordination number (Figs. \ref{fig:fig6}e and f) but slightly overpredicts the viscosity $\eta_r$ as compared to the actual values (Figs. \ref{fig:fig6}d). 
Finally, for the dense limit considered in this study, $\phi =$ 0.80, the actual simulation data (ground truth) itself is noisy, given its vicinity to jamming packing fraction $\phi_J^\mu$. Figures~\ref{fig:fig6} (g-i) show a comparison between actual and validation predictions. It is striking to note that even with the fluctuations in the actual data sets (ground truth), the validation predictions and actual data show excellent agreement. As shown in Figs. \ref{fig:fig6}g–i, although particle pressure and coordination number evolve smoothly, the validation results exhibit greater scatter than the ground truth.  This may arise from the influence of the shared loss function in the multitask regression framework, where (1) discrepancies in the scale of target values can affect optimization as the viscosity and particle pressure values are logarithmically scaled and (2) fluctuations in the viscosity component can impact the overall model prediction, leading to variability in the predicted values for particle pressure and coordination number. Despite minor deviations, the results demonstrate highly accurate interpolations of three key suspension properties, ranging from rheology to microstructure, providing a straightforward tool for understanding dense suspensions.

\section{Concluding Remarks}
\label{Sec:Conclusion}
In this study, we developed microstructure-informed multitask learning models based on Deep Graph Convolutional Neural Networks to predict key rheological and microstructural properties of dense suspensions at each snapshot per stress, including relative viscosity $\eta_r$, particle pressure $\Pi$, and frictional coordination number $Z_\mu$, by encoding interparticle interactions as graph-based representations. 
Our primary goal was to utilize minimal input features, deliberately excluding processing conditions such as $\phi$, $\sigma$, and interparticle forces, thereby allowing the model to identify the hidden features directly from structural topology. As a consequence, the developed models successfully captured the non-linear behavior of semi-dilute to dense suspensions across a wide range of stress conditions by leveraging only geometric features derived from particle configurations. 
%
This approach can effectively be utilized as a mesoscale surrogate to infer underlying patterns and established links between microstructural and macroscopic properties, even for conditions close to jamming, where traditional simulations face challenges due to the computation of force networks and stress tensors during inference.
%
%
%
Future studies might integrate graph neural networks with physics-informed learning to improve extrapolation capabilities. 
Overall, this work establishes a transferable, data-driven strategy for predicting suspension rheology, reducing overall reliance on time-consuming simulations and difficult experiments. Beyond suspensions, the framework holds promise for a broader class of particulate and soft matter systems where structure–property relationships play a central role. This includes network-based systems, such as gels, which can be investigated to provide further insights into phenomena such as shear thinning and yielding. This capability is critical for optimizing material design and enabling predictive modeling in challenging experimental scenarios where bulk properties result from complex multiscale interactions among their components. Moreover, the reliance of this method on physical features potentially broadens its application for real-time characterization and control of suspension flows in an experimental environment.  

\paragraph*{Acknowledgments}
All of this work made use of the High-Performance Computing Resource in the Core Facility for Advanced Research Computing at Case Western Reserve University.
A. S. acknowledges Case Western Reserve University for start-up funding.

\bibliography{dst}
\bibliographystyle{apsrev4-1}
\end{document}